\documentclass[pra,twocolumn,nofootinbib,floatfix,10pt]{revtex4-2}
\usepackage{textcomp,mathcomp}
\usepackage{amsmath}
\usepackage{amssymb}
\usepackage{wasysym}
\usepackage{graphicx}
\usepackage{color,soul}
\usepackage{physics}
\usepackage{siunitx}
\usepackage{dsfont}
\usepackage{float}
\usepackage[english]{babel}
\usepackage{blindtext}
\usepackage[english,nomargin,inline,marginclue,draft]{fixme}
\pdfpageheight\paperheight
\pdfpagewidth\paperwidth

\usepackage[colorlinks,linkcolor=blue,anchorcolor=blue,citecolor=blue,urlcolor=blue]{hyperref}

\fxusetheme{colorsig}
\FXRegisterAuthor{cg}{acg}{CG}  
\FXRegisterAuthor{th}{ath}{\color{blue}TH}  
\FXRegisterAuthor{ib}{aib}{\color{red}IB} 
\FXRegisterAuthor{sh}{ash}{\color{cyan}SH} 
\FXRegisterAuthor{db}{adb}{\color{green}DB} 
\FXRegisterAuthor{ps}{aps}{PS}
\makeatletter
\renewcommand*\FXLayoutInline[3]{%
  {\@fxuseface{inline}\ignorespaces{\color{fx#1}[#3: #2]}}}
\makeatother

\long\def\symbolfootnote[#1]#2{\begingroup%
\def\thefootnote{\fnsymbol{footnote}}\footnotetext[#1]{#2}\endgroup}

\def\nobreakbefore{%
  \relax\ifvmode\else
    \ifhmode
      \ifdim\lastskip > 0pt\relax
        \unskip\nobreakspace
      \else 
        \nobreakspace
      \fi
    \fi
  \fi
}
\let\oldcite\cite
\renewcommand\cite{\nobreakbefore\oldcite}

\begin{document}
\title{Observation of Moiré Time Crystal in Floquet-driven Rydberg Atomic Gases}

\author{Shuai Shi$^{1,4,5\textcolor{blue}{\star}}$}
\author{Dong-Yang Zhu$^{2,3,\textcolor{blue}{\star}}$}
\author{Yu Yang$^{1,4,5,\textcolor{blue}{\star}}$}
\author{Chu-Rong Pan$^{1,4,5,\textcolor{blue}{\star}}$}
\author{Jing-Wen Tang$^{1,4,5}$}
\author{Ya-Peng Zhang$^{1,4,5}$}
\author{Yan-Li Zhou$^{1,4,5}$}
\author{Wei-Tao Liu$^{1,4,5,\textcolor{blue}{\P}}$}
\author{Li-Hua Zhang$^{2,3,\textcolor{blue}{\S}}$}
\author{Bang Liu$^{2,3,\textcolor{blue}{\ddagger}}$}
\author{Dong-Sheng Ding$^{2,3,\textcolor{blue}{\dagger}}$}

\affiliation{$^1$College of Science, National University of Defense Technology, Changsha 410073,China.}
\affiliation{$^2$Laboratory of Quantum Information, University of Science and Technology of China, Hefei, Anhui 230026, China.}
\affiliation{$^3$Anhui Province Key Laboratory of Quantum Network, University of Science and Technology of China, Hefei 230026, China.}
\affiliation{$^4$Interdisciplinary Center for Quantum Information, National University of Defense Technology, Changsha 410073, China.}
\affiliation{$^5$Hunan Research Center of the Basic Discipline for Physical States, National University of Defense Technology, Changsha 410073, China.}

\date{\today}
\symbolfootnote[1]{S.S., D.Y.Z, Y.Y., and C.R.P. contribute equally to this work.}
\symbolfootnote[5]{wtliu@nudt.edu.cn}
\symbolfootnote[4]{zlhphys@ustc.edu.cn}
\symbolfootnote[3]{lb2016wu@ustc.edu.cn}
\symbolfootnote[2]{dds@ustc.edu.cn}

\maketitle

\textbf{A Moiré time crystal is a non-equilibrium quantum phase emerging from the coherent interference of two distinct frequencies, at least one being the intrinsic oscillation of a symmetry-broken time crystal. Its hallmark is an ultra-long beat period, reflecting a time-domain mapping of the Moiré fringes that arise from mismatched spatial lattices. However, to date, no experimental realization of such a Moiré time crystal has been reported. In this work, by applying a bichromatic driving field with two distinct frequencies, we demonstrate that the interplay between long-range Rydberg interactions and dissipation gives rise to a unique comb-like Moiré pattern characterized by a beat-note comb, which superimposes subharmonic periodicity and fundamental frequencies. This Moiré pattern formed by two mismatched drives is staggered in the spectrum as the frequency of one driver changes. We experimentally map the phase diagram of the system and identify a robust region where the Moiré temporal order persists against perturbations in laser detuning. The reported Moiré time crystal not only provides a controllable platform for exploring emergent slow-fast dynamics and synthetic space-time symmetries but also opens avenues for engineering complex temporal order in driven quantum many-body systems.}

\section*{INTRODUCTION}
A Moiré crystal in the spatial domain refers to the long-wavelength superlattice formed by overlaying two periodic lattices with a slight mismatch or twist angle, arising from geometric interference between the constituent lattices \cite{bistritzer2011moire,dai_twisted_2016,cao2018unconventional,lou_tunable_2022,lai_moire_2025}. The physical properties of Moiré crystals are remarkably rich: they host flat electronic bands that suppress kinetic energy and enhance electron-electron correlations, leading to emergent phenomena such as unconventional superconductivity, correlated insulators, orbital magnetism, and topological phases \cite{cao2018correlated,sharpe2019emergent,yankowitz_tuning_2019,andrei2020graphene,li2021lattice,kezilebieke_moire-enabled_2022,nuckolls2023quantum,zheng_superconductivity_2024}. The Moiré periodicity can be continuously tuned by adjusting the twist angle or lattice mismatch, providing a highly controllable platform for band structure engineering, while also supporting exciton localization and interlayer exciton condensation, which makes Moiré crystals attractive for quantum optics and ultracold atoms \cite{tran2019evidence,fu_optical_2020,zhang2021van,wang2020localization,tang_-chip_2022,yu_moire_2023,luan2023reconfigurable,meng2023atomic,wang2024three,saadi_tailoring_2025}. A natural temporal analogue is the Moiré time crystal (MTC), where two periodic drives produce a slow beat-note evolution \cite{zou2024momentumflatbandsuperluminalpropagation,dong2025extremely,liang2026atomic}, which is distinct from the discrete time crystal (DTC) with a single driving \cite{wilczek2012quantum,sacha2017time,autti_observation_2018,else2020discrete,Mark_Lyubarov2022,zaletel2023colloquium,yousefjani_discrete_2025}. Its observation in strongly interacting quantum many-body systems with tunable long-range interactions is still lacking. 

The strong interactions inherent in Rydberg atoms make driven-dissipative Rydberg ensembles an ideal testbed for exploring a variety of non-equilibrium phenomena, ranging from self-organization and non-equilibrium phase transitions \cite{lee2012collective,carr2013nonequilibrium,helmrich2020signatures,ding2019Phase,ding2022enhanced,wadenpfuhl2023emergence,ding2023ergodicity} to the emergence of dissipative time crystals \cite{wu2023observation,liu2024bifurcation,jiao2025observation,jiao2025photoionization}, higher-order and fractional-order DTCs \cite{liu2024higher}, and time quasicrystals~\cite{Logarithmically2018,quasiperiodic2018,Quasi-Floquet2023,zhu2025observation}. In such systems, a periodic external drive can induce subharmonic responses, providing a clear manifestation of broken discrete time-translation symmetry~\cite{PhysRevLett.120.140401}. Therefore, Rydberg atoms offer a unique platform to extend the study of MTCs into the strongly interacting regime, where the interplay between long-range interactions and bichromatic driving can give rise to complex temporal order beyond single driving. In addition, the frequency‑matching conditions that govern time crystals under multi‑frequency driving have yet to be investigated.

\begin{figure*}[t] 
    \centering \includegraphics[width=0.95\textwidth]{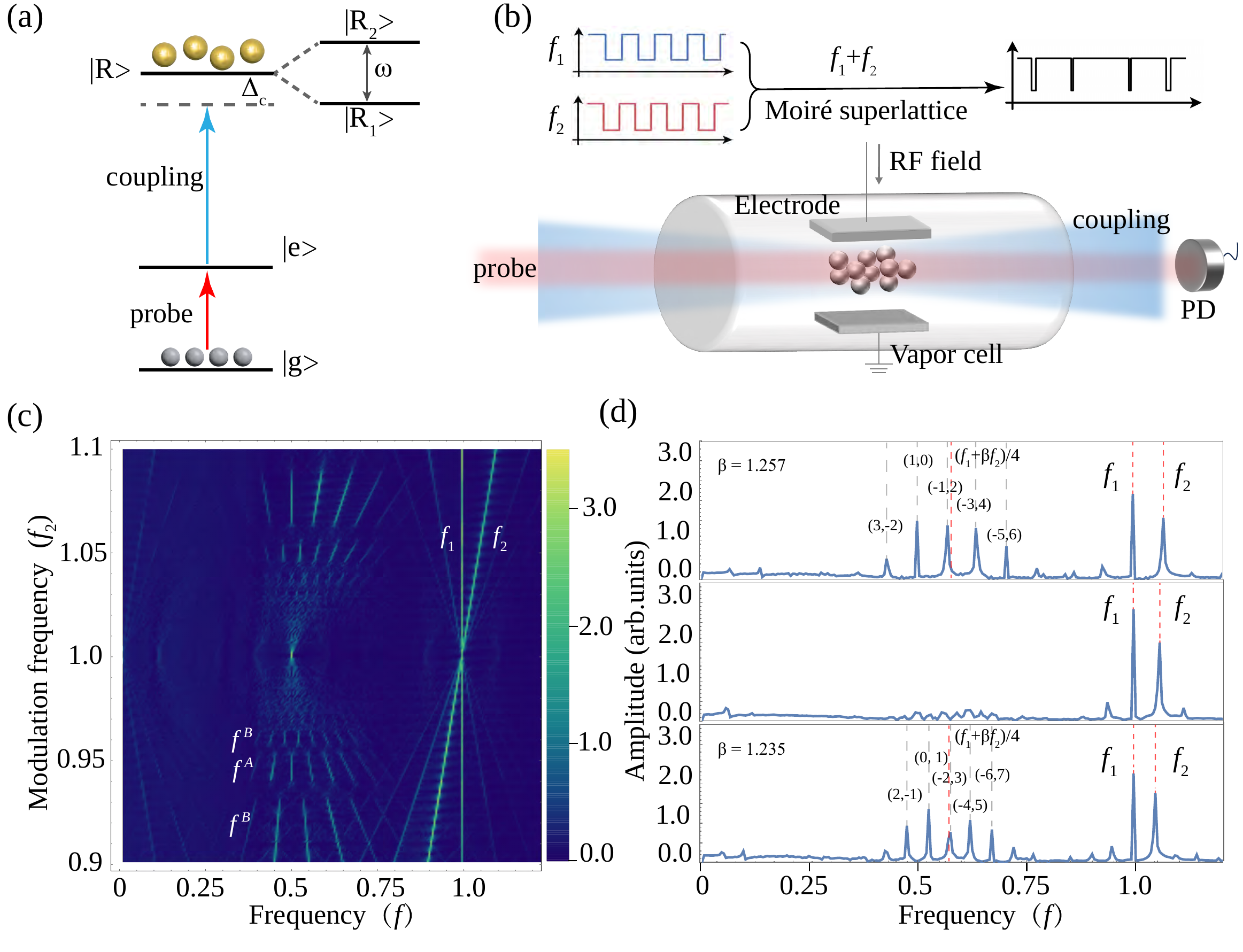} 
    \caption{\textbf{Experimental diagram and theoretical calculations} (a) Schematic energy level diagram for a two-photon excitation scheme. The probe laser drives the transition from the ground state $|g\rangle$ to the intermediate state $|e\rangle$, while the coupling laser excites the transition from $|e\rangle$ to the Rydberg state $|R\rangle$ with a detuning $\Delta_c$. The radio-frequency (RF) field driving induces the Rydberg state $|R\rangle$ into Floquet sidebands $|R_1\rangle$ and $|R_2\rangle$, with an energy spacing of $\omega$. (b) Simplified experimental setup. The probe and coupling lasers propagate counter-propagating and pass through the vapor cell. The bichromatic RF field is applied to the atoms via two parallel electrode plates built inside the cell, forming time superlattice drive. (c) Calculated phase diagram of Fourier spectra of the Rydberg population $\rho_{nR_1}$ under the bichromatic field Floquet driving. (d) Fourier spectra at $f_2$ = 1.07$f_1$ (up), 1.06$f_1$ (middle), and 1.05$f_1$ (down). The comb pattern of Moiré time crystals switches between subharmonic series of the form $m f_1$/2 + $n f_2/2$, each comb tooth is marked as ($m$, $n$). The red dashed line denotes the weighted average of several intermediate frequency combs, with a frequency given by $(f_1 + \beta f_2)/4$, where $\beta = 1.257$ for $f_2 = 1.07f_1$ and $\beta = 1.235$ for $f_2 = 1.05f_1$.} 
    \label{fig.1} 
\end{figure*}

In this work, we experimentally realize a MTC in a driven-dissipative ensemble of strongly interacting Rydberg atoms. By applying a bichromatic driving field, we observe the emergence of a beat-note comb in the Fourier spectra. This comb-like MTC exhibits a staggered structure versus frequency difference due to the frequency matching conditions of discrete time-translation $\mathbb{Z}{_2}$-symmetry, directly reflecting the spontaneous breaking of time‑translation symmetry driven by the bichromatic driving competition. We systematically map the phase diagram of the system by varying the laser detuning, revealing a robust feature where the Moiré temporal order persists against perturbations. These observations establish the Rydberg atomic platform as a versatile testbed for exploring time-translation symmetry breaking under competition and open new avenues for engineering complex temporal structures in driven quantum many-body systems. 

\begin{figure*}[t] 
    \centering \includegraphics[width=0.95\textwidth]{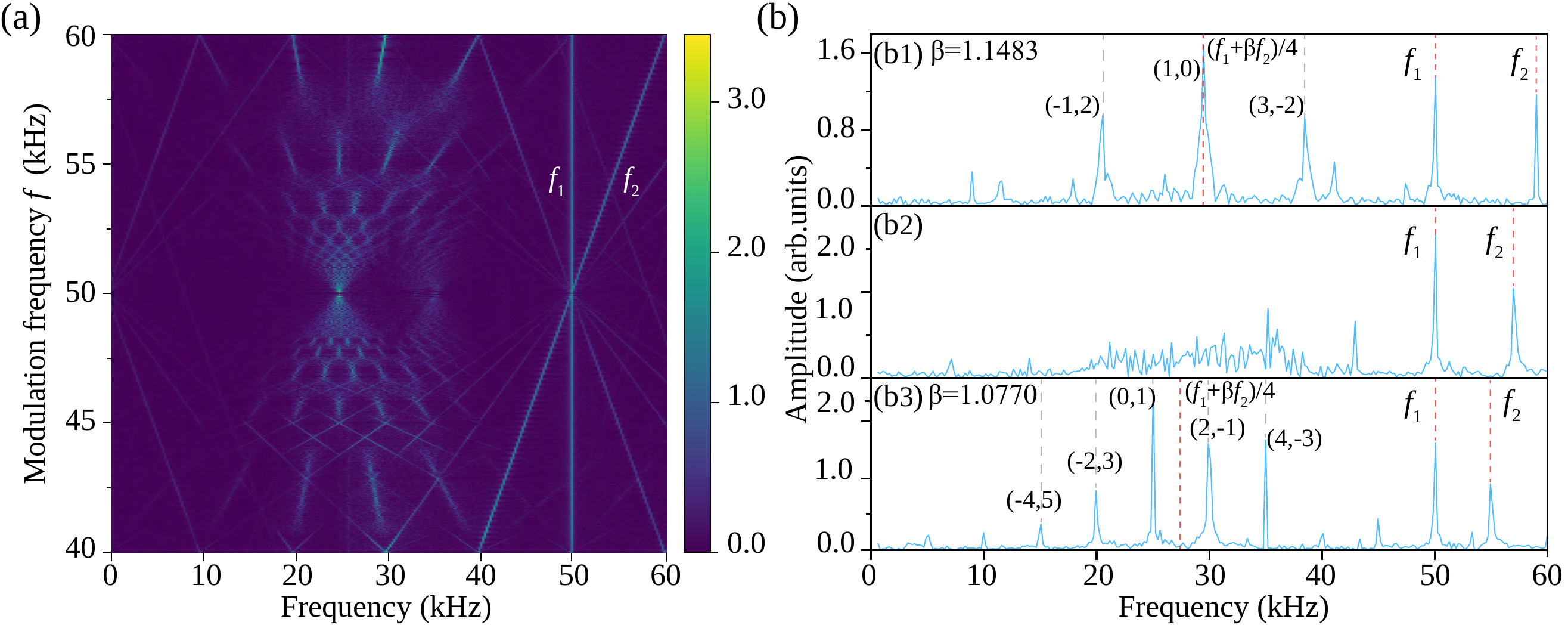} 
    \caption{\textbf{Measured Moiré time crystal} (a) Color map showing the probe light transmission while scanning the $f_2$-field. In the experiment, the RF $f_1$-field is set to a frequency of 13 MHz, an amplitude $U_1 = 250$ mVpp, and a modulation frequency $f_1 = 50$ kHz. The RF $f_2$-field has a frequency of 13 MHz and an amplitude $U_2 = 260$ mVpp. The color map is obtained by scanning the modulation frequency $f_2$ of the $f_2$-field from 40 kHz to 60 kHz. The color bar represents the transmission intensity. (b) Examples of the Fourier spectra of the system response at different modulation frequencies. (b1) When $f_2 = 59$ kHz, the Fourier spectrum shows a Moiré pattern of a frequency comb centered at $m f_1 / 2 + n f_2/2$. Panel (b2) shows the Fourier spectrum when $f_2 = 57.05$ kHz, where the subharmonic frequency comb in the system response disappears. (b3) shows the Fourier spectrum when $f_2 = 55.02$ kHz, where the system response exhibits a Moiré pattern of a frequency comb centered at $m f_1/2 + n f_2/2$. The red dashed lines in the figure mark the positions of the two driving frequencies $f_1$ and $f_2$. The gray dashed lines labeled with $(m, n)$ in (b1) and (b3) indicate the frequencies of the subharmonic frequency comb.} 
    \label{fig.2} 
\end{figure*}

\section*{Physical model}
To characterize the emergence of MTCs, we consider a driven-dissipative ensemble of three-level Rydberg atoms, where each atom consists of a ground state $\left| g \right\rangle$ and two Rydberg excited states $\left| R_1 \right\rangle$ and $\left| R_2 \right\rangle$. The coherent dynamics of the system are described by a Hamiltonian under dual-frequency driving \cite{wu2023observation,liu2024higher},
\begin{equation}
\begin{aligned}
    \hat{H}(t) & =\frac{1}{2}\sum_{i}\left(\Omega_{1}\sigma_{i}^{gR_1}+\Omega_{2}\sigma_{i}^{gR_2}+h.c.\right)\\
    &-\sum_{i}\left[(\Delta_{f_1}(t)+\Delta_{f_2}(t))(n_{i}^{R_1}+n_{i}^{R_2})+\delta n_{i}^{R_2}\right] \\
    &+\sum_{i\neq j}V_{ij}\bigg[n_{i}^{R_1}n_{j}^{R_2}+\frac{1}{2}(n_{i}^{R_1}n_{j}^{R_1}+n_{i}^{R_2}n_{j}^{R_2})\bigg].
\end{aligned}
\label{hamiltonian}
\end{equation}
Here, $\sigma_{i}^{gr}$ ($r=R_1,R_2$) denotes the transition operator between $\left| g \right\rangle$ and $\left| r \right\rangle$ on site $i$, while $n_i^{R_1}$ and $n_i^{R_2}$ are the corresponding population operators. The dual-frequency modulation is encoded in the time-dependent detunings $\Delta_{f_1}(t)$ and $\Delta_{f_2}(t)$, and $\delta$ characterizes the energy offset between the two Rydberg levels. The interaction between atoms at positions $\mathbf{r}_i$ and $\mathbf{r}_j$ is taken to be of van der Waals form, $V_{ij}=C_6/|\mathbf{r}_i-\mathbf{r}_j|^6$. Dissipation is incorporated through Lindblad channels describing spontaneous decay from $\left| r \right\rangle$ to $\left| g \right\rangle$, i.e., $\mathcal{L}_r=(\gamma_r/2)\sum_i(\hat{\sigma}_i^{rg}\hat{\rho}\hat{\sigma}_i^{gr}-\{\hat{n}_i^r,\hat{\rho}\})$ for $r=R_1,R_2$. Within the mean-field approximation, the density matrix evolves according to $\partial_t\hat{\rho}=i[\hat{H},\hat{\rho}]+\mathcal{L}_{R_1}[\hat{\rho}]+\mathcal{L}_{R_2}[\hat{\rho}]$, from which we determine the dynamical phase diagram by analyzing the behavior of $\rho_{R_1R_1}(t)$. 

Under single-frequency driving, i.e., when only $f_1$ or $f_2$ is applied, the system is periodically forced and responds at the same period as the drive. However, the nonlinear interactions between Rydberg atoms prevent the atomic evolution from fully following the period of the driving field, thus inducing the breaking of discrete time-translation symmetry $\mathbb{Z}{_m}$, giving rise to a subharmonic response at $f=f_d/m$ (with integer $m>1$ and $f_d=f_1$ or $f_2$), which signals the formation of a DTC \cite{liu2024higher}. When the two driving frequencies are simultaneously applied and $f_1$ and $f_2$ are slightly detuned from each other, the system response contains two subharmonic components at $f_1/2$ and $f_2/2$.

We fix the driving frequency $f_1$ and sweep the other driving frequency $f_2$ from $0.9f_1$ to $1.1f_1$. The resulting spectral response phase diagram of the system is shown in Fig. \ref{fig.1}(c). At $f_2 = 1.07f_1$, the spectral response given by Fig. \ref{fig.1}(d1) exhibits a Moiré pattern of a subharmonic frequency comb centered at $f^{A}=m f_1$/2 + $n f_2/2$ (with odd $m$ and even $n$), with a spacing equal to $f_2 - f_1$. When $f_2 = 1.06f_1$, the subharmonic frequency comb vanishes, and only the driving frequencies $f_1$ and $f_2$ remain, indicating that time-translational symmetry is not broken, see Fig. \ref{fig.1}(d2). At $f_2 = 1.05f_1$, the spectral response [Fig. \ref{fig.1}(d3)] shows a frequency comb centered at $f^{B}=m f_1$/2+ $n f_2/2$ (with even $m$ and odd $n$), again with spacing $f_2 - f_1$. As $f_2$ is varied, a Moiré pattern of the frequency comb is staggered between the dominant subharmonic frequency $f_1/2$ and $f_2/2$, separated by an intermediate comb-free region. This staggered variation is spaced by the frequency difference of $(f_1-f_2)/2$ for adjacent comb teeth.

\begin{figure*}[t] 
    \centering \includegraphics[width=\textwidth]{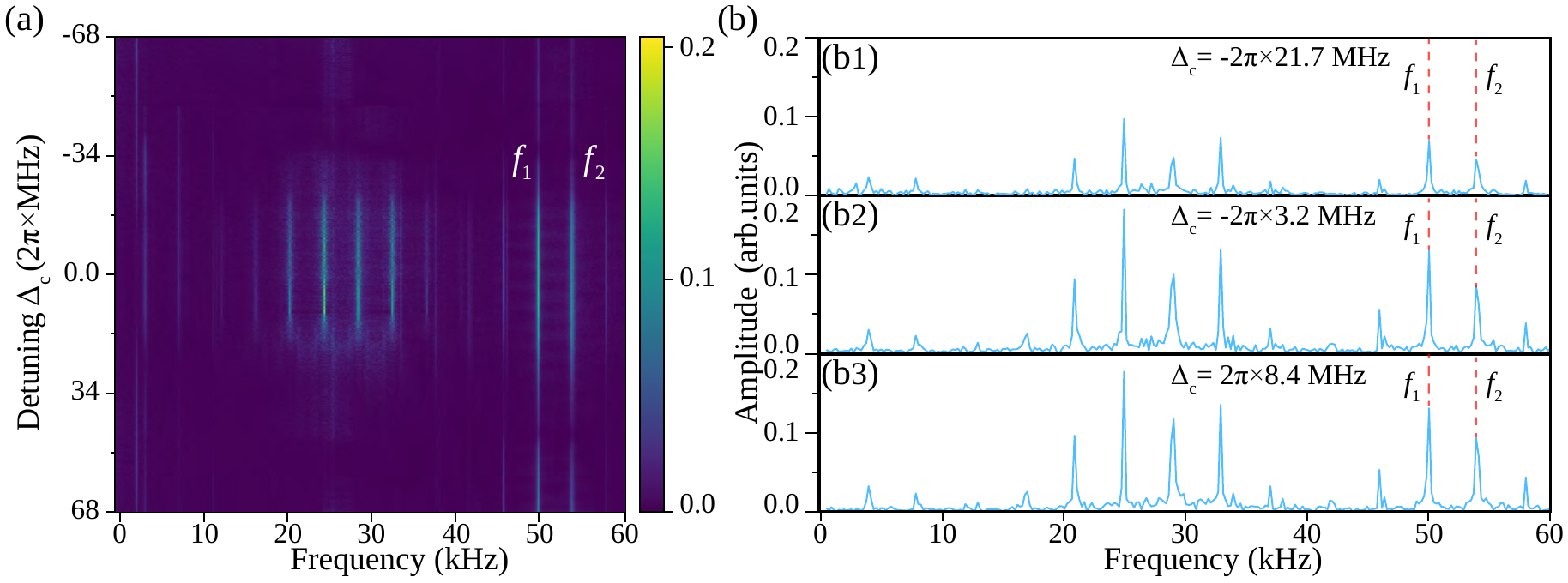} 
    \caption{\textbf{Robustness of the Moiré time crystal pattern versus $\Delta_c$.} The frequencies of the $f_1$-field and the $f_2$-field are set to 13 MHz. The amplitude of the $f_1$-field is $U_1 = 280$ mV, and that of the $f_2$-field is $U_2 = 282$ mV. The modulation frequency of the $f_1$-field is $f_1 = 50$ kHz, and that of the $f_2$-field is set to $f_2 = 54$ kHz. The system response exhibits a Moiré pattern of a subharmonic frequency comb dominated by $f_2/2$. (a) The phase diagram of the system response is obtained by scanning the coupling laser detuning $\Delta_c$ from $\Delta_c = -2\pi \times 68$ MHz to $\Delta_c = 2\pi \times 68$ MHz. The color bar represents the transmission intensity. (b) Fourier spectra of the probe transmission at different coupling detunings $\Delta_c$. (b1) $\Delta_c = -2\pi \times 21.7$ MHz; (b2) $\Delta_c = -2\pi \times 3.2$ MHz; (b3) $\Delta_c = 2\pi \times 8.4$ MHz. The red dashed lines mark the driving frequencies $f_1$ and $f_2$.} 
    \label{fig.3} 
\end{figure*}

To quantitatively characterize the comb-like structure of MTCs, we define the spectral response function as a superposition of two distinct subharmonic families:
\begin{equation}
S(f) = \sum_{m,n} \left[ A_{m,n} \, \delta\!\left(f -f^{A}\right) + B_{m,n} \, \delta\!\left(f - f^{B}\right) \right],
\tag{2}
\end{equation}
where $m,n \in \mathbb{Z}$ are integers, $A_{m,n}$ and $B_{m,n}$ denote the spectral weights (intensities) of the respective comb teeth, and $\delta$ is the Dirac delta function. The first term contains the subharmonic series centered at $f_1/2$ with sidebands spaced by $f_2 - f_1$, the second term describes another series centered at $f_2/2$ with the same spacing. The weights for these two subharmonic series are defined as $S_A = \sum_{m,n} A_{m,n}$, $S_B=\sum_{m,n}B_{m,n}$. This decomposition directly captures the staggered switching behavior observed in Fig.~\ref{fig.1}(d). When $S_A \gg S_B$, the comb is dominated by the first series; when $S_B \gg S_A$, the second series prevails; and when both $S_A$ and $S_B$ are suppressed, the system enters the comb-free region. This depends on whether the intensity-weighted center frequency of the set of comb teeth in the MTC lies closer to half of the intensity-weighted center frequency of the two driving channels $f_1$ and $f_2$. When a comb center resides sufficiently close to the half-frequency mark of $\bar{f}$, the system satisfies the frequency-matching condition for a 2:1 parametric resonance, allowing the drive to efficiently pump energy into a symmetry-broken, period-doubled orbit. When neither set is sufficiently close, the system enters the comb-free regime.

The center frequency of the entire frequency comb can be obtained through the intensity-weighted average of all frequency components:
\begin{equation}
\bar{f} = \frac{\sum_{m,n} f^A A_{m,n} + \sum_{m,n} f^B B_{m,n}}{S_A + S_B}
\tag{3}
\end{equation}
This formula determines the effective center of the Moiré frequency comb from the spectral weights. According to our theoretical calculations, the red dashed line in Figs.~\ref{fig.1}(c) and (d) marks the weighted average of several intermediate frequency combs, with the comb center fixed at $\bar{f} \approx (f_1+\beta f_2)/4$. Here $\beta$ represents the response coefficient from the difference between two driving frequencies, with $\beta = 1.257$ when $f_2 = 1.07f_1$, and $\beta = 1.235$ when $f_2 = 1.05f_1$. This result indicates that the spectral energy of the MTC is the linear combination of the two driving frequencies $f_1$ and $f_2$. 

\section*{Results}
\subsection{Measured MTC and staggered effect}
To experimentally investigate the emergence of MTCs, we prepare a strongly interacting Rydberg atomic ensemble in a room-temperature rubidium vapor cell. The Rydberg states are excited through a two-photon  electromagnetically induced transparency (EIT) scheme consisting of counter-propagating 780 nm probe and 480 nm coupling lasers \cite{mohapatra2007coherent}. The corresponding energy-level configuration and experimental setup are illustrated in Figs.~\ref{fig.1}(a) and (b), with additional details provided in the Methods section.

\begin{figure*}[t] 
    \centering \includegraphics[width=\textwidth]{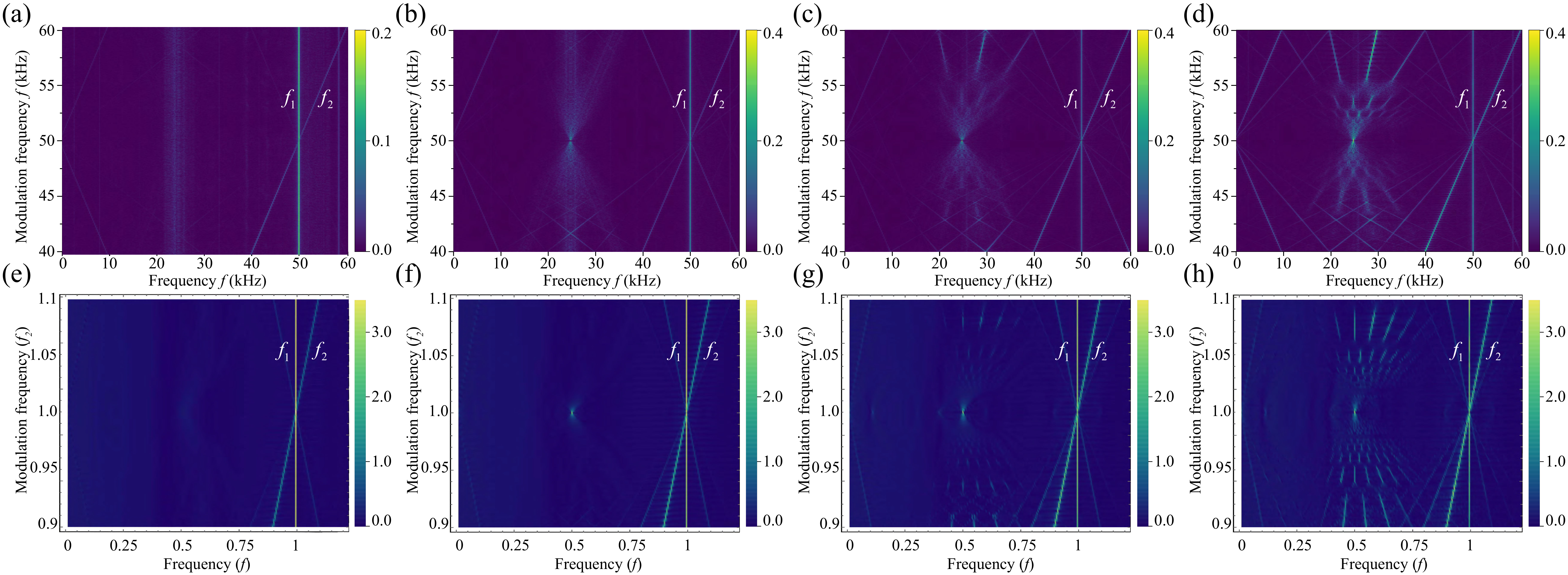} 
    \caption{\textbf{Measured and theoretical phase diagrams versus $f_2$-field intensity.} The frequency of the $f_1$-field is set to 13 MHz, with a modulation frequency $f_1 = 50$ kHz and amplitude $U_1 = 280$ mV kept constant. The frequency of the $f_2$-field is set to 13 MHz, with the amplitude $U_2$ gradually increased, scanning the modulation frequency $f_2$ from 40 kHz to 60 kHz to get phase diagrams of the system response under different RF field strengths. (a) $f_2$ field strength $U_2 = 50$ mV; (b) $f_2$ field strength $U_2 = 140$ mV; (c) $f_2$ field strength $U_2 = 260$ mV; (d) $f_2$ field strength $U_2 = 280$ mV. Panels (e)–(h) show the theoretical phase diagrams for driving strengths of $f_2$ equal to 0.2, 0.3, 0.43, and 0.49, respectively.} 
    \label{fig.4} 
\end{figure*}

In the experiment, we apply two independently controlled radio-frequency (RF) fields to the Rydberg ensemble through parallel electrodes embedded inside the vapor cell. Each RF field is modulated by a square waveform with frequencies of $f_1$ and $f_2$, respectively. When applied separately, either driving field induces a subharmonic response of the interacting Rydberg system, characterized by a frequency component $f_i/2$, demonstrating the formation of a $\mathbb{Z}{_2}$-DTC phase. We then simultaneously apply both RF drives, fixing $f_1$ = 50 kHz while continuously tuning $f_2$ from 40 to 60 kHz. The time-domain response signal of the transmitted probe was recorded and subsequently Fourier transformed to obtain the frequency-domain response spectrum, as shown in Fig.~\ref{fig.2}(a). 

The measured spectrum in Fig.~\ref{fig.2}(a)  reveals the formation of a Moiré temporal structure characterized by a staggered subharmonic frequency comb. In addition to the two fundamental driving frequencies $f_1$ and $f_2$, a series of subharmonic frequency components emerge and evolve continuously as $f_2$ is tuned. These responses are organized into two distinct frequency-comb families centered around the half-frequency components $f_1/2$ and $f_2/2$, corresponding to two independent symmetry-breaking channels induced by the bichromatic driving. As the frequency mismatch $\Delta_f = f_2 - f_1$ between the two driving fields is varied, the dominant subharmonic response switches periodically between the two channels, resulting in a staggered distribution of the frequency comb in the spectrum. This switching behavior reflects the nontrivial breaking of time-translation symmetry beyond the direct combination of the conventional single-frequency-driven DTC.
 
Representative Fourier spectra extracted from three characteristic regions of Fig.~\ref{fig.2}(a) are presented in Figs.~\ref{fig.2}(b1)–(b3). At $f_2$ = 59 kHz (Fig.~\ref{fig.2}(b1)), the system breaks into the driving channel $ f_1$, the Moiré pattern composed of multiple subharmonic peaks centered at $f^{A}=m f_1$/2 + $n f_2/2$ (with odd $m$ and even $n$). When $f_2$ = 57.05 kHz (Fig.~\ref{fig.2}(b2)), the time-translational symmetry is not broken and the system is in an intermediate comb-free region. The subharmonic frequency comb vanishes, and only the driving frequency $ f_1$ and $ f_2$ remain. At $f_2$ = 55.02 kHz (Fig.~\ref{fig.2}(b3)), the system breaks into the driving channel $ f_2$,  the Moiré pattern composed of multiple subharmonic peaks centered at $f^{B}=m f_1$/2 + $n f_2/2$ (with even $m$ and odd $n$).

The experimentally measured Fourier spectrum clearly exhibits the main peaks at the driving frequencies $f_1$ and $f_2$, as well as the subharmonic responses of the system. In contrast to the DTC driven by a single frequency, the system under dual-frequency driving exhibits richer subharmonic responses, manifested as a Moiré interference pattern. The measured Fourier spectrum clearly reveals two alternating subharmonic response peaks, independently located at the half-frequency components $f_1/2$ and $f_2/2$. A pronounced comb-free region is also observed between the two patterns, indicating competition between them. The typical Fourier spectrum of these three characteristic regions is presented in Figs.~\ref{fig.2}(b1)–(b3). The appearance of these two main peak patterns signifies that the system oscillates in the time domain with periods twice those of the respective driving fields. 

\begin{figure*}[t] 
    \centering \includegraphics[width=\textwidth]{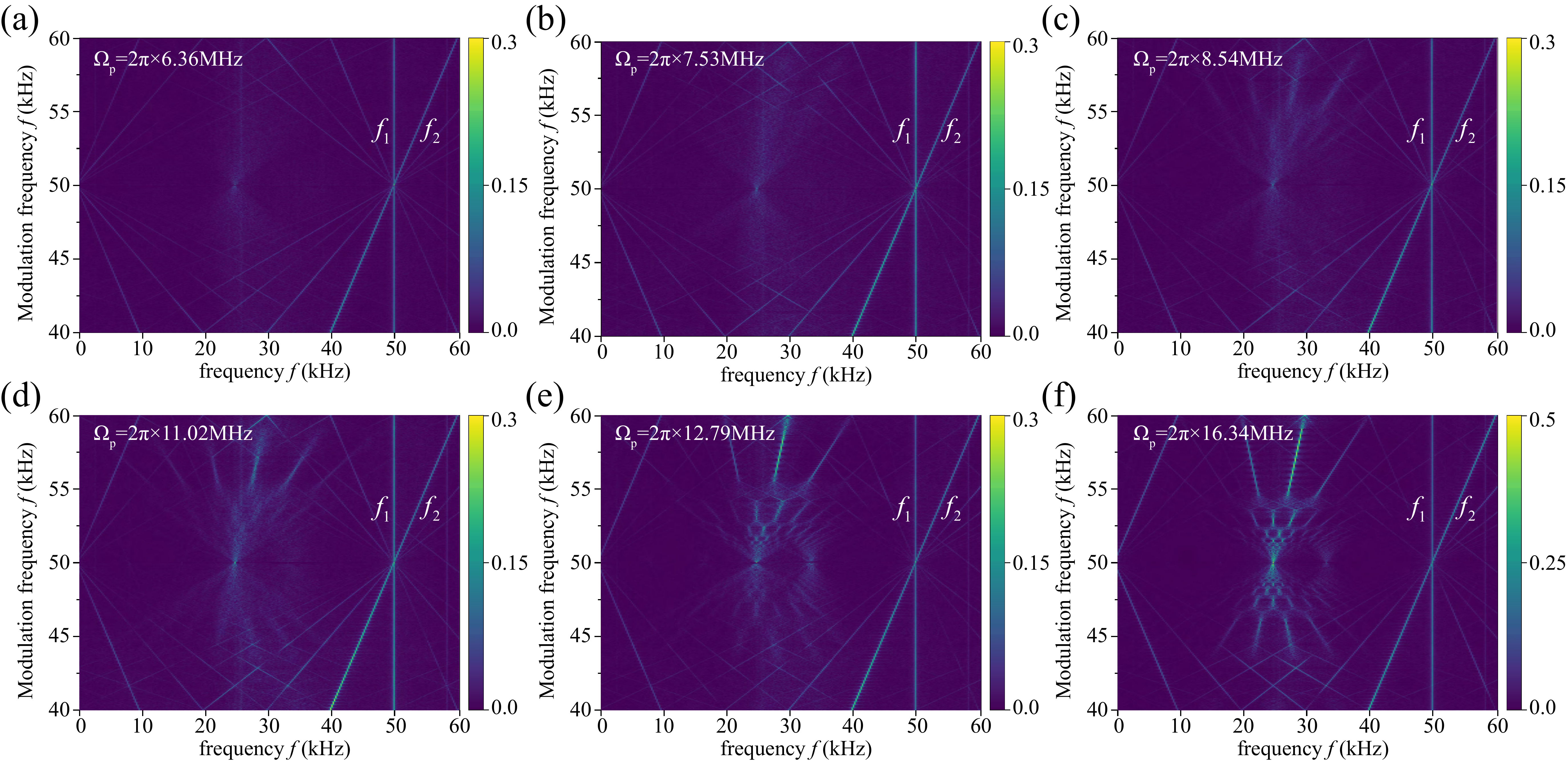} 
    \caption{\textbf{Phase diagrams obtained by varying the probe Rabi frequency $\Omega_p$.} The frequencies of the $f_1$-field and the $f_2$-field are set to 13 MHz, with modulation frequency $f_1 = 50$ kHz, amplitudes $U_1 = 370$ mV and $U_2 = 380$ mV. Phase diagrams of the system response are obtained by scanning $f_2$ from 40 kHz to 60 kHz at different probe Rabi frequencies. Panels (a)--(f) correspond to progressively increasing probe Rabi frequencies, as indicated in the upper-left corners: (a) $\Omega_p = 2\pi \times 6.36$ MHz, (b) $\Omega_p = 2\pi \times 7.53$ MHz, (c) $\Omega_p = 2\pi \times 8.54$ MHz, (d) $\Omega_p = 2\pi \times 11.02$ MHz, (e) $\Omega_p = 2\pi \times 12.79$ MHz, and (f) $\Omega_p = 2\pi \times 16.34$ MHz.} 
    \label{fig.5} 
\end{figure*}

Surrounding each main peak, the spectrum unfolds into a series of equally spaced sidebands, with a spacing exactly equal to the difference between the two driving frequencies $\Delta_f = f_2 - f_1$. This gives rise to overlapping comb-like structures, arising from the coupling of two temporal lattice oscillations via the nonlinear Rydberg interaction and producing a frequency-domain beat-frequency modulation analogous to a spatial Moiré pattern. The result reveals the emergence of a time-ordered state beyond the single-frequency-driven DTC, arising from the competition of two time crystals and demonstrating emergent rich features in the Fourier spectrum of the MTC. In addition, when considering the case where $f_1$ = 40 kHz and $f_2$ = 60 kHz, the dual-RF driving produces a longer-period MTC arising from the mixing of the two subharmonic components $f_1/2$ and $f_2/2$. In this case, the competition is replaced by a cooperative contribution from both time crystals to form the longer-period MTC, as detailed in the supplementary materials.

\subsection{Robust of MTC}

The robustness of the MTC against laser detuning perturbations was investigated by varying the coupling field detuning $\Delta_c$ while keeping all other parameters fixed. The modulation frequencies of the two RF fields are set to $f_1$ = 50 kHz and $f_2$ = 54 kHz.The system breaks into driving channel $f_2$ and the corresponding Moiré frequency comb is centered around $f_2/2$. The measured phase diagram is presented in Fig.~\ref{fig.3}(a). By scanning the coupling field detuning $\Delta_c$ from $-2\pi\times68$ to $+2\pi\times68$ MHz, the phase diagram comprehensively displays the global behavior of the system response under different detuning conditions. To examine the stability of the MTC pattern against perturbations in $\Delta_c$, we measured Fourier spectra at three specific detuning values in the phase diagram, $\Delta_c = -2\pi\times21.7,-2\pi\times3.2 ,2\pi\times8.4$ MHz. The corresponding Fourier spectra, presented in Figs.~\ref{fig.3}(b1)–(b3), all exhibit clear Moiré frequency comb patterns. The MTC pattern stably exists within the detuning range of 43 MHz, which is close to the EIT spectrum linewidth. Despite the wide variation of $\Delta_c$, the frequency comb pattern remains stable without appreciable degradation or disappearance, demonstrating the robustness of the MTC against perturbations in the coupling field detuning.

\subsection{MTC under asymmetric drives}

To investigate the characteristics of MTC phases under asymmetric driving conditions, we measured the phase diagrams at various $f_2$-field intensity while keeping the $f_1$-field intensity constant. During the measurements, the frequency, amplitude, and modulation frequency of the $f_1$-field were maintained at 13 MHz, 280 mV, and 50 kHz, respectively. The $f_2$-field frequency was kept constant at 13 MHz throughout. The phase diagram of MTC can be obtained by measuring the variation of the Fourier spectrum of the probe signal as a function of the modulation frequency of the $f_2$ field. By gradually increasing the intensity of $f_2$-field to the same level as $f_1$-field, we obtained the phase diagrams under different field intensity, as shown in Figs.~\ref{fig.4}(a)–(d). As the $f_2$ field intensity increases, the moiré structure in the phase diagram progressively becomes more distinct. This indicates that the MTC arises from the competition between two periodic drivings, and the asymmetry in the intensity of the two driving fields affects the contrast of the moiré structure. Only when the amplitudes of the two drivings are equal can the two time crystal phases fully compete, thereby generating a clear MTC phase. Moreover, we also numerically simulated the phase diagrams under different $f_2$ field intensities, as shown in Figs.~\ref{fig.4}(e)–(h), and the theoretical results are in excellent agreement with the experimental observations.

\subsection{Many-body origin of MTCs}

The moiré structure observed in this experiment arises from the competition between two DTC with distinct periods. DTC emerge in strongly interacting many-body systems under periodic driving. Within such systems, variations in the interaction strength can modify the properties of the DTC.Consequently, this leads to alterations in the structure of the resulting MTC.  To investigate the effect of interaction strength on the MTC phase, we measured the phase diagrams under different probe light intensities. The probe light intensity directly influences the population of Rydberg atoms, which increases with the probe light intensity. Given that the total atomic density remains constant at a fixed vapor cell temperature, the increase in Rydberg population leads to a reduction in the interatomic spacing of Rydberg atoms, thereby enhancing the interactions among them. Experimentally, we held the coupling light parameters constant and measured the phase diagrams for probe Rabi frequencies of 6.36 MHz, 7.53 MHz, 8.54 MHz, 11.02 MHz, 12.79 MHz and 16.34 MHz, respectively, as shown in Figs.~\ref{fig.5}(a)–(f). The experimental results indicate that as the interaction strength increases, the breaking of discrete time-translation symmetry becomes more pronounced. This leads to a gradual decrease in the frequency components in the phase diagram that correspond to the driving frequency, while the MTC phase composed of subharmonic frequency components gradually emerges and becomes enhanced.

\section*{Discussions}
The realization of MTCs in a driven-dissipative Rydberg atomic ensemble marks a significant advance in the study of many-body non-equilibrium quantum phases. Unlike conventional DTCs that arise from a single periodic drive and exhibit a single subharmonic response, the MTC emerges from temporal superlattice response, giving rise to a breaking of time-translation symmetry under competition. The observed MTC manifests as a staggered frequency comb in the Fourier spectrum, whose structure is directly tunable by the frequency mismatch between the two driving fields. This switching behavior reflects a competition between two symmetry-breaking channels, reminiscent of the geometric frustration or lattice mismatch effects in spatial Moiré systems. 

Our experimental observations are in good agreement with the theoretical phase diagram derived from the mean-field master equation, which predicts the alternating dominance of two subharmonic families centered at $f_1$/2 and $f_2$/2, separated by a comb-free region. From a broader perspective, the MTC provides a controllable platform for exploring emergent non-equilibrium dynamics with multi-frequency driving. The ability to engineer complex temporal structures opens new avenues for studying non-equilibrium phases that are not accessible in single-drive systems. The possible extension to higher-order or fractional MTCs merit further theoretical and experimental investigation. Our work establishes a foundation for these future studies and highlights the driven-dissipative Rydberg atomic platform as a versatile testbed for exploring the rich phenomenology of quantum many-body systems.

\section*{Summary}
In summary, we have experimentally observed MTCs in a strongly interacting, driven-dissipative Rydberg atomic gas. By applying a bichromatic radio-frequency driving field with a varying frequency difference, we observed the emergence of a staggered subharmonic frequency comb in the system’s response spectrum. This comb structure arises from the spontaneous breaking of time-translation symmetry under competition and exhibits two distinct subharmonic families centered at half of each driving frequency, with sidebands spaced by the frequency difference between the two drives. We systematically mapped the phase diagram of the system over a wide range of laser detuning and identified a robust Moiré temporal order. 

When the two driving fields couple to the atoms with equal strength, the two subharmonic families contribute symmetrically to the Moiré frequency comb. As a result, the overall comb center falls exactly at the midpoint of the two half‑drive frequencies, i.e. $(f_1+ f_2)/4$. In this balanced case the asymmetry parameter $\beta$ naturally equals unity. Physically, $\beta$ thus quantifies the relative effective weights of the two channels. Any deviation from equal coupling—due to unequal intensities, detuning, or nonlinear mixing—breaks the symmetry and shifts the comb center away from that simple average, yielding $\beta \neq 1$.

The observed phenomena are well captured by our theoretical model based on a mean-field treatment of the Rydberg interaction and dissipation, which reproduces the key features of the spectral response. Our findings not only extend the concept of Moiré patterns from the spatial domain to the temporal domain but also establish a new class of non-equilibrium quantum phases—the MTC—that bridges the physics of time crystals, Floquet engineering, and strongly interacting Rydberg systems. This work paves the way for future explorations of synthetic temporal lattices, emergent slow-fast dynamics, and the engineering of complex quantum many-body states with tailored temporal correlations.

\section*{Methods}

\subsection*{Master Equations}
We can calculate the dynamical evolution of the system via the Lindblad master equation approach. Owing to the thermal motion of the atoms, correlations between atoms can be neglected, allowing us to adopt the mean-field approximation. From the Hamiltonian of the system, we obtain the master equation:
\begin{equation}
\begin{aligned}
\frac{\partial}{\partial t} \rho_{R_1 R_1} & =i \frac{\Omega}{2}\left(\rho_{g R_1}-\rho_{R_1 g}\right)-\gamma \rho_{R_1 R_1}, \\
\frac{\partial}{\partial t} \rho_{R_2 R_2} & =i \frac{\Omega}{2}\left(\rho_{g R_2}-\rho_{R_2 g}\right)-\gamma \rho_{R_2 R_2}, \\
\frac{\partial}{\partial t} \rho_{g R_1} & =i \frac{\Omega}{2}\left(\rho_{R_1 R_1}+\rho_{R_2 R_1}-\rho_{g g}\right) \\ 
& +i\left((\Delta_{f_1}(t)+\Delta_{f_2}(t) )-V_{\rm{MF}}+i \frac{\gamma}{2}\right) \rho_{g R_1}, \\
\frac{\partial}{\partial t} \rho_{g R_2} & =i \frac{\Omega}{2}\left(\rho_{R_2 R_2}+\rho_{R_1 R_2}-\rho_{g g}\right)\\
& +i\left((\Delta_{f_1}(t)+\Delta_{f_2}(t) )+\delta-V_{\rm{MF}}+i \frac{\gamma}{2}\right) \rho_{g R_2}, \\
\frac{\partial}{\partial t} \rho_{R_1 R_2} & =i \frac{\Omega}{2}\left(\rho_{g R_2}-\rho_{R_1 g}\right)-i\left(\delta-i \gamma\right) \rho_{R_1 R_2},
\end{aligned}
\end{equation}
where $V_{\rm{MF}} = V(\rho_{R_1 R_1}+\rho_{R_2 R_2})$ is the mean field shift, and we set the effective Rabi frequency $\Omega_1=\Omega_2=\Omega$. Here, we treat a simplified model by using same interaction $V_{ij}=V$ by ignoring the difference between different sublevels of Rydberg atoms. The term $\Delta_{f_1}(t)+\Delta_{f_2}(t)$ represents the total time-dependent detuning induced by a dual-frequency driving field. In our numerical simulation, this total detuning is implemented as a square-wave modulation with two distinct frequency components, which mimics the experimental scheme of dynamically modulating the laser frequency. By solving the equations above, we can obtain the time response of the system and can also obtain the Fourier spectrum via discrete Fourier transformation as shown in Figs.~\ref{fig.1}(c) and (d). In the calculations, all populations and coherences are initially set to zero, i.e., the system starts from the ground state. 

\subsection*{Experimental setup}
The experimental system is based on rubidium-87 atoms in a 6 cm long, 4 cm diameter vapor cell. We employ a two-photon EIT scheme to prepare Rydberg atoms and monitor the probe transmission to measure the population distribution of the Rydberg atoms, thereby characterizing the dynamical evolution of the system. The energy level structure of the rubidium atoms used in the experiment is shown in Fig.~\ref{fig.1}(a). A 780 nm probe laser drives the transition from $|5S_{1/2}, F = 2\rangle$ to $|5P_{3/2}, F = 3\rangle$ with Rabi frequency $\Omega_p$. A 480 nm coupling laser drives the transition from $|5P_{3/2}, F = 3\rangle$ to the Rydberg state $|60D_{5/2}\rangle$ with Rabi frequency $\Omega_c$ and detuning $\Delta_c$. Under the driving of the RF field, the Rydberg levels generate a series of Floquet sidebands, denoted as $|R_1\rangle$ and $|R_2\rangle$, with an energy spacing of $\omega$.

In the experimental setup, a 780 nm external cavity diode laser (ECDL) is locked using saturated absorption spectroscopy (SAS) as a frequency reference. The beam is split into probe and reference beams, which pass through the vapor cell in parallel, with a $1/e^2$ beam waist radius of approximately 500 $\mu$m and a power of 400 $\mu$W. A frequency-doubled laser (TA-SHG pro) produces a 480 nm coupling beam that counter-propagates with the probe beam inside the cell, with a $1/e^2$ beam waist radius of about 700 $\mu$m and a power of 1.2 W, as shown in Fig. \ref{fig.1}(b). The Rabi frequencies of the probe and coupling lasers are $\Omega_p = 2\pi \times 35$ MHz and $\Omega_c = 2\pi \times 4.3$ MHz, respectively. These configurations produce electromagnetically induced transparency, making the probe beam transparent. Finally, the probe and reference beams are differentially amplified by a balanced photodetector, and the probe transmission is recorded with an oscilloscope. We use an arbitrary function generator (AFG, RIGOL DG 902 pro) to produce the periodically modulated RF electric field required in the experiment, which is applied via a pair of parallel electrode plates built into the vapor cell, with a length of 4 cm, a width of 2.5 cm, and a separation of 2.5 cm. By connecting the coupling laser, oscilloscope, and AFG to a computer for coordinated control, the external parameters can be rapidly scanned to obtain the phase diagram of the system response.

\section*{Acknowledgements}
We acknowledge funding from the National Natural Science Foundation of China (Grant Nos. T2495253), the National Key R and D Program of China (Grant No. 2022YFA1404002).

\hspace*{\fill}

\section*{Data Availability}
All experimental data used in this study are available from the corresponding author upon request.

\section*{Author contributions statement}
D.-S.D. and S.S conceived the idea and supported this research. D.Y.Z, Y. Y., and S. S. conducted the physical experiments. D.-S.D. and B.L. conducted the theoretical calculations. The manuscript was written by All authors. All authors contributed to discussions regarding the results and the analysis contained in the manuscript.

\section*{Competing interests}
The authors declare no competing interests.

\bibliography{ref}

\end{document}


\title{Supplementary information: Observation of Moiré Time Crystal in Floquet-driven Rydberg Atomic Gases}

\author{Shuai Shi$^{1,4,5\textcolor{blue}{\star}}$}
\author{Dong-Yang Zhu$^{2,3,\textcolor{blue}{\star}}$}
\author{Yu Yang$^{1,4,5,\textcolor{blue}{\star}}$}
\author{Chu-Rong Pan$^{1,4,5,\textcolor{blue}{\star}}$}
\author{Jing-Wen Tang$^{1,4,5}$}
\author{Ya-Peng Zhang$^{1,4,5}$}
\author{Yan-Li Zhou$^{1,4,5}$}
\author{Wei-Tao Liu$^{1,4,5,\textcolor{blue}{\P}}$}
\author{Li-Hua Zhang$^{2,3,\textcolor{blue}{\S}}$}
\author{Bang Liu$^{2,3,\textcolor{blue}{\ddagger}}$}
\author{Dong-Sheng Ding$^{2,3,\textcolor{blue}{\dagger}}$}

\affiliation{$^1$College of Science, National University of Defense Technology, Changsha 410073,China.}
\affiliation{$^2$Laboratory of Quantum Information, University of Science and Technology of China, Hefei, Anhui 230026, China.}
\affiliation{$^3$Anhui Province Key Laboratory of Quantum Network, University of Science and Technology of China, Hefei 230026, China.}
\affiliation{$^4$Interdisciplinary Center for Quantum Information, National University of Defense Technology, Changsha 410073, China.}
\affiliation{$^5$Hunan Research Center of the Basic Discipline for Physical States, National University of Defense Technology, Changsha 410073, China.}

\date{\today}
\symbolfootnote[1]{S.S., D.Y.Z, Y.Y., and C.R.P. contribute equally to this work.}
\symbolfootnote[5]{wtliu@nudt.edu.cn}
\symbolfootnote[4]{zlhphys@ustc.edu.cn}
\symbolfootnote[3]{lb2016wu@ustc.edu.cn}
\symbolfootnote[2]{dds@ustc.edu.cn}

\maketitle

The supplementary material contains the measured 2-DTC for single-tone RF‑$f_1$ or RF‑$f_2$ driving, the Moiré time crystal (MTC) comb pattern under dual-RF loading, the temporal dynamics of the MTC, and the experimentally obtained phase diagram versus RF frequency. 

\subsection*{Measured 2-DTC with RF-$f_1$ and RF-$f_2$ driving}
\begin{figure*}[t] 
    \centering \includegraphics[width=0.95\textwidth]{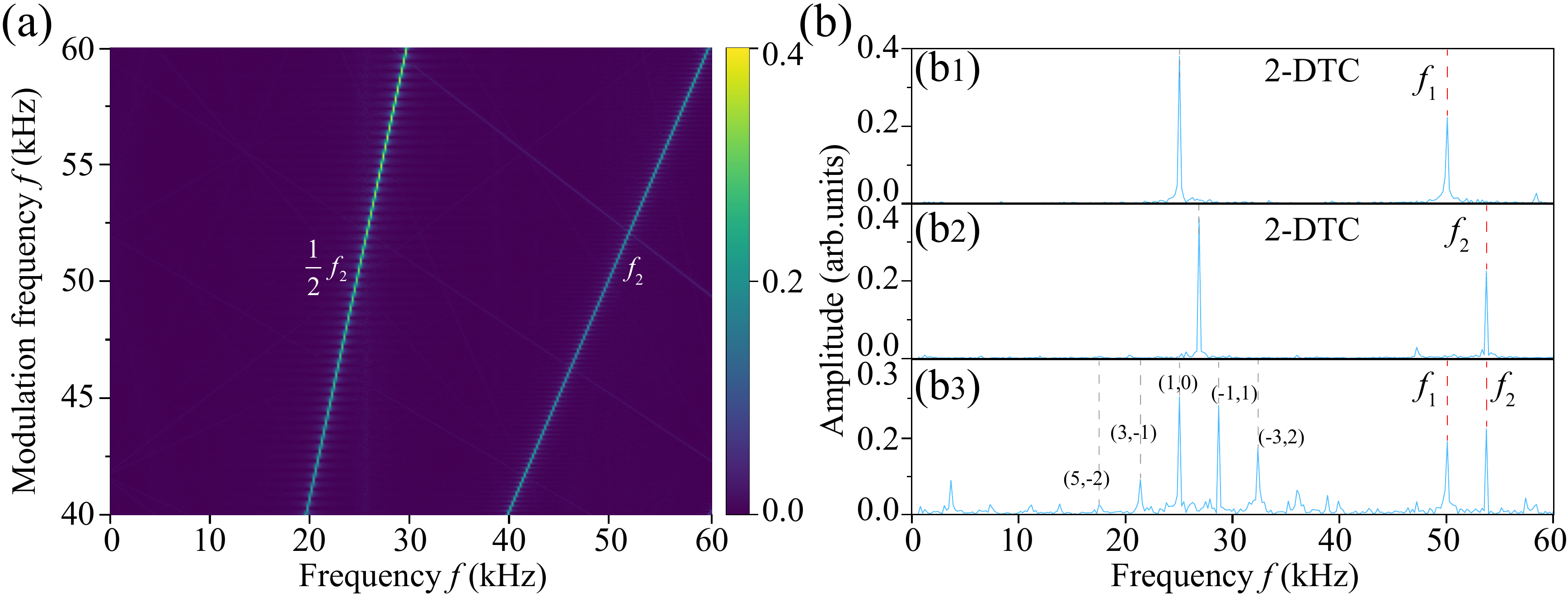} 
    \caption{\textbf{The measured systematic response was characterized under single-frequency ($f_1$ or $f_2$) and dual-frequency (simultaneous $f_1 + f_2$) RF field driving.} (a) Phase diagram of the system response showing a 2-DTC signature, obtained by sweeping $f_2$ from 40 to 60 kHz with the $f_1$ field off ($f_2$ carrier: 13 MHz, $U_2$ = 280 mV). (b) Example Fourier spectra of the system response. (b1) With only the $f_1$-field turned on at $f_1 = 50$ kHz and the $f_2$ field switched off, the system exhibits a 2-DTC. (b2) Under driving by the $f_2$-field alone at $f_2 = 53.7$ kHz, with the $f_1$ field turned off, the system exhibits a 2-DTC. (b3) When both the $f_1$ and $f_2$ fields are simultaneously present, the system exhibits a Moiré frequency comb dominated by a subharmonic family centered at $f_1/2$.} 
    \label{fig.1} 
\end{figure*}

To establish the individual contributions of the two driving fields to the formation of the MTC response, we first characterized the system under each RF modulation separately. As shown in  Fig.~\ref{fig.1}(a), when only the $f_2$-field is applied, scanning $f_2$ from 40 to 60 kHz produces two pronounced spectral branches following $f_2$ and $f_2/2$,demonstrating that the $f_2$-field alone can induce a $\mathbb{Z}{_2}$-DTC response over a broad range of modulation frequencies \cite{liu2024higher}. This behavior is further illustrated by the representative Fourier spectra in Fig.~\ref{fig.1}(b). With only the $f_1$-field applied at $f_1=50$ kHz, a dominant subharmonic peak appears at $f_1/2$ (Fig.~\ref{fig.1}(b1)), whereas driving the system solely with the $f_2$-field at $f_2=53.7$ kHz gives a corresponding peak at $f_2/2$ (Fig.~\ref{fig.1}(b2)). These observations are consistent with the theoretical model, in which nonlinear Rydberg interactions prevent the atomic dynamics from rigidly following the external modulation and stabilize a symmetry-broken dynamical state whose period is twice that of the drive.  

When the two RF fields are applied simultaneously, the two independently accessible period-doubling channels become coupled, and the isolated subharmonic peak evolves into an equally spaced frequency comb described by $f^{A}=m f_1$/2 + $n f_2/2$ (with integers $m$ and $n$), as shown in Fig.~\ref{fig.1}(b3). These measurements therefore support the interpretation that the Moiré frequency comb originates from nonlinear mixing and competition between two single-drive DTC responses, rather than from a simple superposition of the two fundamental driving frequencies.

\subsection*{Time-flow of the Moiré time crystal}
\begin{figure*}[t] 
    \centering \includegraphics[width=0.95\textwidth]{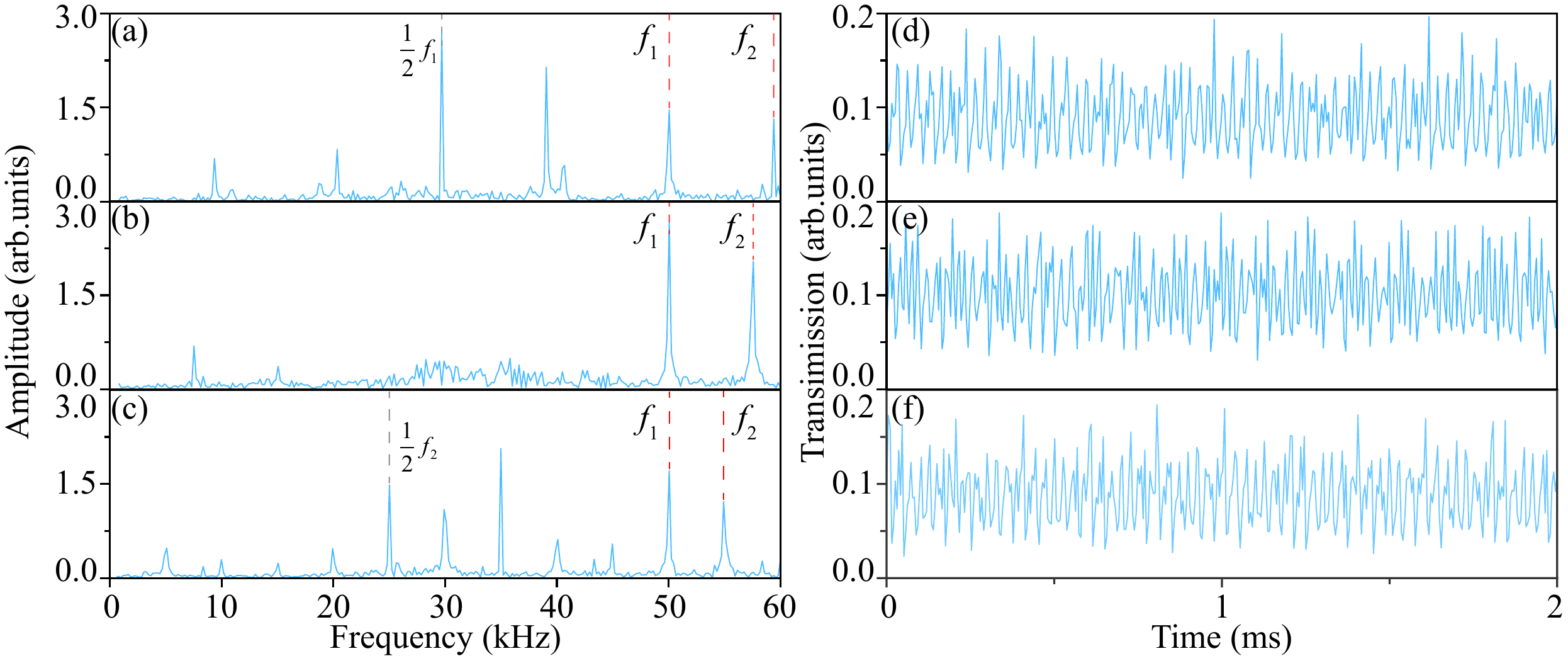} 
    \caption{\textbf{Time-flow of the system response under different states.} The carrier frequencies of both the $f_1$-field and the $f_2$-field are set to 13 MHz, with modulation frequency $f_1 = 50$ kHz, amplitudes $U_1 = 320$ mV and $U_2 = 340$ mV. (a) At $f_2 = 55$ kHz, the system displays a Moiré frequency comb from the mixing of $f_1/2$ and $f_2-f_1$. (b) At $f_2 = 57.5$ kHz, the system transitions into a comb-free state. (c) At $f_2 = 59.4$ kHz, the system exhibits a Moiré frequency comb originating from the mixing of $f_2/2$ and $f_2-f_1$. (d)--(f) Corresponding time-domain results for the three different states, respectively.} 
    \label{fig.2} 
\end{figure*}

The measured Fourier spectra presented in this paper are obtained by applying a Fourier transform to the temporal oscillations of the probe transmission intensity extracted from the electromagnetically induced transparency transmission spectrum \cite{mohapatra2007coherent}. To provide a comprehensive view of the subharmonic response at the Moiré pattern dominated by either frequency $f_1$ or $f_2$, we present the time-domain and frequency-domain diagrams of the probe transmission spectrum for three typical regions, as shown in Fig.~\ref{fig.2}.

\subsection*{Measured phase diagram versus varying RF field frequency}
\begin{figure*}[t] 
    \centering \includegraphics[width=0.95\textwidth]{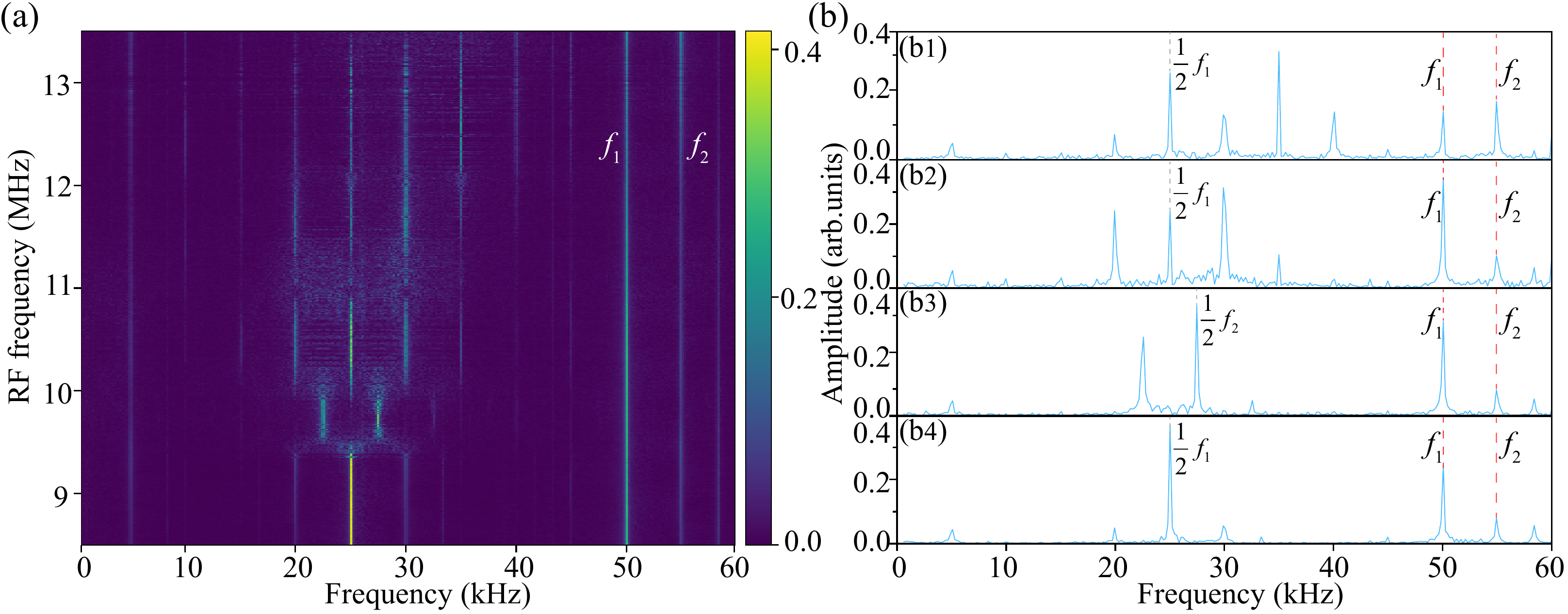} 
    \caption{\textbf{Phase diagram obtained by varying the RF field frequency.} The $f_1$-field is set to a carrier frequency of 13 MHz, with modulation frequency $f_1 = 50$ kHz and amplitude $U_1 = 320$ mV. The $f_2$-field has a modulation frequency $f_2 = 55$ kHz and amplitude $U_2 = 300$ mV. (a) Phase diagram of the system response obtained by scanning the carrier frequency of the $f_2$-field from 8.5 MHz to 13.5 MHz. (b) Example Fourier spectra of the system response at different carrier frequencies. (b1) At an $f_2$-field carrier frequency of 13.02 MHz, the system displays a Moiré frequency comb dominated by the mixing of $f_1/2$ and $f_2-f_1$. (b2) At a carrier frequency of 11.5 MHz, the system displays a Moiré frequency comb the mixing of $f_1/2$ and $f_2-f_1$ with different comb tooth strength. (b3) At a carrier frequency of 9.7 MHz, the system enters a state dominated by the mixing of $f_2/2$ and $f_2-f_1$. (b4) When a carrier frequency is 9 MHz, the system exhibits a subharmonic response of $f_1$, entering a 2-DTC state.} 
    \label{fig.3} 
\end{figure*}

Since the MTC originates from the competition between two subharmonic components, the center frequency of the Moiré frequency comb is determined by the dominant subharmonic component. To examine how the MTC frequency comb evolves with the competition between two 2-DTC with distinct periods, we can adjust the carrier frequency of one driving field to control the result of the competition between the two subharmonic components. In the experiment, the carrier frequency, amplitude, and modulation frequency of the $f_1$ field were held at 13 MHz, 320 mV, and 50 kHz, respectively, while the amplitude and modulation frequency of the $f_2$ field were set to 300 mV and 55 kHz. By measuring the Fourier spectrum of the probe transmission signal as a function of the carrier frequency of the $f_2$ field, we obtained the MTC phase diagram shown in Fig.~\ref{fig.3}(a), where the carrier frequency of the $f_2$ field was gradually increased from 8.5 MHz to 13.5 MHz. 

The measured phase diagram reveals that, as the carrier frequency of the $f_2$ field varies, the dominant frequency of the Moiré frequency comb switches back and forth between $f_1$/2 and $f_2$/2 dominant mixing structures, reflecting the evolving competition between the two subharmonic components. Figure.~\ref{fig.3}(b) presents the corresponding Fourier spectra at four distinct carrier frequencies of the $f_2$ field. Figures.~\ref{fig.3} (b1-b2) show the moiré frequency combs at carrier frequencies of 13.02 MHz and 11.5 MHz, respectively, where the subharmonic component of the $f_1$ driving field dominates. When the $f_2$ carrier frequency is reduced to 9.7 MHz, as shown in Fig.~\ref{fig.3}(b3), the dominant frequency of the moiré frequency comb shifts to $f_2$/2, indicating that the subharmonic component of the $f_2$ driving field becomes dominant. Finally, at a carrier frequency of 9 MHz, the system responds only with a subharmonic response to the $f_1$ driving field; consequently, Fig.~\ref{fig.3}(b4) exhibits the 2-DTC signal solely at the $f_1$/2 frequency. These experimental results demonstrate the tunable effect of the carrier frequency on the competition between subharmonic components and its influence on the Moiré frequency comb.

\subsection*{Measured Fourier spectra under dual-RF driving}
\begin{figure*}[t] 
    \centering \includegraphics[width=0.95\textwidth]{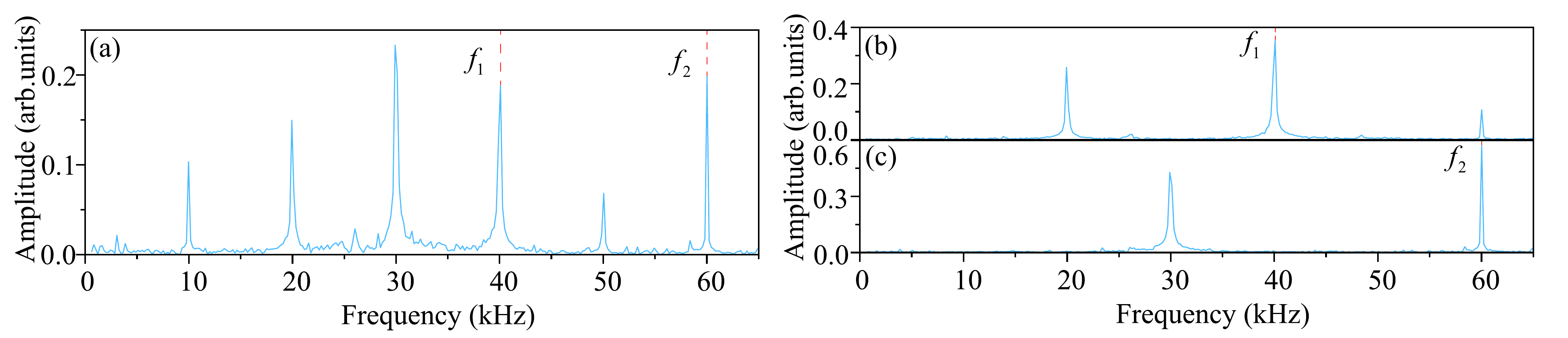} 
    \caption{\textbf{Measured Fourier spectra under dual-RF loading at $f_1=40$ kHz and $f_2=60$ kHz}. Both RF-fields are set to a carrier frequency of 13 MHz, with modulation frequencies $f_1 = 40$ kHz and $f_2 = 60$ kHz and amplitudes $U_1 = 380$ mV and $U_2 = 340$ mV. (a) Measured subharmonic response in the Fourier spectra under dual-RF driving. The response at 10 kHz corresponds to the mixing of $f_2/2$ and $f_1/2$, revealing a longer periodicity MTC. (b) With only the $f_1$-field applied at $f_1 = 40$ kHz, the system exhibits a 2‑DTC response with a frequency of $f_1/2$. (c) Driven by the $f_2$-field alone at $f_2 = 60$ kHz, the system exhibits a 2‑DTC response with a frequency of $f_2/2$.}
    \label{fig.4} 
\end{figure*}
We further consider the scenario where $f_1$ = 40 kHz and $f_2$ = 60 kHz. Under dual-RF driving with both fields at a 13 MHz carrier and amplitudes $U_1 = 380~\text{mV}$ and $U_2 = 340~\text{mV}$, the measured Fourier spectrum reveals a pronounced subharmonic peak at 10 kHz [Fig.~\ref{fig.4}(a)]. This response is particularly noteworthy because it cannot be attributed to the discussed MTC mechanisms in main text—namely, the mixing of $f_2/2$ with $f_2-f_1$ or $f_1/2$ with $f_2-f_1$. Instead, the 10 kHz peak arises from the difference-frequency mixing between $f_2/2 = 30~\text{kHz}$ and $f_1/2 = 20~\text{kHz}$, indicating a longer-period MTC. To further validate this interpretation, we also performed control measurements with only one field active: panel (b) in Fig.~\ref{fig.4} shows that the $f_1$-field alone at 40 kHz produces a 2-DTC response at $f_1/2 = 20~\text{kHz}$, while panel (c) in Fig.~\ref{fig.4} demonstrates that the $f_2$-field alone at 60 kHz yields a 2-DTC at $f_2/2 = 30~\text{kHz}$. Neither of these single-field configurations generates the 10 kHz peak, confirming that the longer-period MTC observed in Fig.~\ref{fig.4}(a) is a genuine consequence of the interplay between the two subharmonic responses under dual-RF driving.

\bibliography{ref}